\documentclass[aps,pre,preprint]{revtex4-1}

\usepackage{epsfig}
\usepackage{latexsym}
\usepackage{amsmath}
\usepackage{amssymb}
\usepackage{amsfonts}
\usepackage{graphicx}
\usepackage{wrapfig}
\usepackage[left]{lineno}
\usepackage{url}
\usepackage{color,bm}

\usepackage[normalem]{ulem}
\usepackage{color}

\definecolor{red}{rgb}{1,0,0}
\definecolor{green}{rgb}{0,1,0}
\definecolor{blue}{rgb}{0,0,1}

\begin{document}

\title{Experimental and numerical study of a second-order transition in the behavior of confined self-propelled particles}

\author{E. Barone}
\affiliation{Universidad de Buenos Aires, Facultad de Ciencias Exactas y Naturales, Departamento de Física. Buenos Aires, Argentina}%

\author{G. A. Patterson}
\email{gpatters@itba.edu.ar}
\affiliation{Instituto Tecnol\'ogico de Buenos Aires (ITBA), CONICET, Iguazú 341, 1437 C. A. de Buenos Aires, Argentina}%

\date{\today}

\begin{abstract}

In this study, we conduct experimental investigations on the behavior of confined self-propelled particles within a circular arena, employing small commercial robots capable of locomotion, communication, and information processing. These robots execute circular trajectories, which can be clockwise or counterclockwise, based on two internal states. Using a majority-based stochastic decision algorithm, each robot can reverse its direction based on the states of two neighboring robots. By manipulating a control parameter governing the interaction, the system exhibits a transition-from a state where all robots rotate randomly to one where they rotate uniformly in the same direction. Moreover, this transition significantly impacts the trajectories of the robots. To extend our findings to larger systems, we introduce a mathematical model enabling characterization of the order transition type and the resulting trajectories. Our results reveal a second-order transition from active Brownian to chiral motion. Lastly, we analyze the particle density within the arena, examining how it varies concerning system size and the control parameter.

\end{abstract}

\maketitle

\section{Introduction}

Active matter comprises systems of particles that consume energy to propel themselves or perform work \cite{ramaswamy2010mechanics,fodor2018statistical}. One of the possible characteristics that these systems may exhibit is random behavior, exemplified by the trajectories of particles. These disturbances may be caused by environmental influences, internal stochastic processes related to the locomotion mechanism, or even decision-making arising from social interactions \cite{romanczuk2012active}. An active Brownian particle (ABP) is a type of particle model that describes the diffusive behavior, similar to that observed in some living organisms, and is characterized by the random change in the orientation of motion. This trait is adopted by various living beings as a strategy for environmental exploration, either to seek food or to distance themselves from potential threats \cite{berg2004ecoli,cates2012diffusive,poon2013clarkia}. One type of active particle that has received much attention in recent years is chiral active particles (CAP). Chirality is not only a property of the geometry of objects but also of the trajectories that self-propelled particles undertake when the symmetry of the direction of motion is broken. There are various examples in nature of particles exhibiting chiral behavior. For instance, \textit{E. coli} bacteria \cite{diluzio2005escherichia,lauga2006swimming} and spermatozoa \cite{riedel2005selforganized,friedrich2007chemotaxis} exhibit chiral behavior when moving near a glass surface. There are also artificial systems, such as microswimmers, where the specific geometry of the particles generates chiral trajectories \cite{kumemel2013circular}.

The simplest model to describe the behavior of a chiral active particle is based on specifying a translational velocity and an angular rotation \cite{liebchen2022chiral}. Based on this definition, many studies have focused on exploring repulsion interactions \cite{cates2015motility,sese2022microscopic}, polar alignments \cite{liebchen2017collective,ventejou2021susceptibility}, and synchronization \cite{levis2019activity,ventejou2021susceptibility}. Furthermore, there are studies that have investigated how the parameters governing the reorientation of particles determine the type of emerging trajectory of a particle confined in a channel \cite{vanTeeffelen2008dynamics}. 

In this study, we investigate an experimental system of confined particles that can be programmed to execute circular motion, communicate with neighbors, and process received information. While previous works considered interactions aimed at aligning the direction of motion \cite{ginelli2016physics,baglietto2018flocking}, in this study, a majority-based stochastic interaction governs the rotation direction of the particles. The experiments were conducted using a commercial robot named Kilobot \cite{rubenstein2012kilobot}. This type of agents has previously been employed to investigate various phenomena such as pattern formation \cite{rubenstein2014programmable}, decision-making \cite{valentini2016collective}, food searching \cite{fontllenas2018quality}, \cite{patterson2022bistability}, and decentralized learning \cite{patterson2022bistability}.

The goal of this work is to characterize the emerging behavior of the system due to the stochastic interaction. In Sec.~\ref{sec:exp}, we present experimental results demonstrating a shift in the system's behavior by varying a control parameter in the decision-making mechanism. In Sec.~\ref{sec:num}, we introduce a mathematical model for the robots, enabling numerous simulations to characterize the scaling behavior and emerging trajectories, as well as analyze the role of arena boundaries in the particle density. Finally, in Sec.~\ref{sec:con}, we discuss our conclusions.
 
\section{Experimental results}
\label{sec:exp}

The Kilobots are of $3.3\ \mathrm{cm}$ of diameter, $3.4\ \mathrm{cm}$ of tall and they stand on three legs: one front and two rear. Each robot features two vibrators which are controlled independently by an internal microprocessor and allows the robot to rotate in one direction or the other around one of its rear legs at a speed of $0.5\ \mathrm{rad/s}$. Moreover, the robots have the ability to communicate with others within a distance of approximately $10\ \mathrm{cm}$. They can transmit information, such as the current rotation direction, at a rate of $2$ messages per second. This communication, which operates unidirectionally, enables the microprocessor to update internal variables or robot motion based on the received information. 

\begin{figure}
\includegraphics{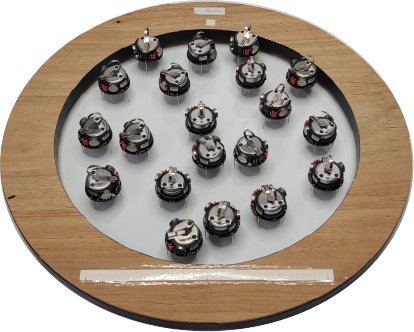}
\caption{\label{fig:setup}Experimental setup. 20 Kilobots were positioned within a circular arena with a radius of $15\ \mathrm{cm}$. The labels used to track the orientation of the particles are visible in the picture.}
\end{figure}

The experimental setup consisted of a circular arena with a radius of $15\ \mathrm{cm}$ mounted on a white melamine board (Fig.\ref{fig:setup}). We placed 20 robots inside the arena whose trajectories were recorded by an overhead video camera at a rate of 1 frame per second. As shown in the figure, we attached two circular markers on the robots to measure their absolute orientation and direction of rotation.

The motion of each robot was generated by clockwise or counterclockwise rotations. The direction of rotation was determined by an internal variable $\sigma_i$, which took a value of $1$ for clockwise or $-1$ for counterclockwise. Each robot updated the value of $\sigma_i$ with a period $T=1\ \mathrm{s}$ following the stochastic rule:
\begin{align}
\nonumber \sigma_i \leftarrow -\sigma_i&\ ,\ \mathrm{with\ probability\ 1\ if}\ \Phi_i < 0\ ,\\
\sigma_i \leftarrow -\sigma_i&\ ,\ \mathrm{with\ probability\ }p\mathrm{\ if}\ \Phi_i \geq 0\ ,\\
\nonumber \sigma_i \leftarrow \sigma_i&\ ,\ \mathrm{with\ probability\ }1-p\mathrm{\ if}\ \Phi_i \geq 0\ ,
\end{align}    
where $\Phi_i = \sigma_i\sum_{j=1}^M \sigma_j$. $M$ is the number of random messages considered, out of all those received by robot $i$ in each period $T$. We arbitrarily chose $M=2$, so $\Phi_i$ could take values of $-2$, $0$, or $2$. While $\Phi_i=-2$ implies that robot $i$ rotates in the opposite direction to the other two robots involved in the interaction, $\Phi_i=2$ determines that all three robots rotate in the same direction. In the case of $\Phi_i=0$, robot $i$ has the same direction of rotation as one of the other robots, and it is considered to have the direction of the majority. The rule states that if the direction is opposite to the majority, with probability $1$, the new rotation state will follow the majority. On the other hand, if the robot rotates in the same direction as the majority, then, with probability $p$, it inversely changes its rotation direction. In the case where no messages have been received during the period $T$, the robot updates its rotation state randomly.

We studied the emergent behavior of the system as a function of the control parameter $p$. For this, we chose eight probability values in the range $p=\left[0.0-0.2 \right]$, and for each of these, we let the system evolve for 60 minutes from random initial conditions (both spatial location and rotation state). Using image analysis, we were able to ascertain $\sigma_i$ for every robot within each frame, enabling us to quantify the system's overall state by employing a conventional order parameter defined as:
\begin{align}
s(t) = & \frac{1}{N}\sum_{i=1}^{N}\sigma_i(t)\ ,
\end{align}
where \(s(t) = 1\) indicates that all robots rotate clockwise, \(s(t) = -1\) counterclockwise, and \(s(t) \approx 0\) suggests a balance between the two directions of rotation.

\begin{figure}
\includegraphics{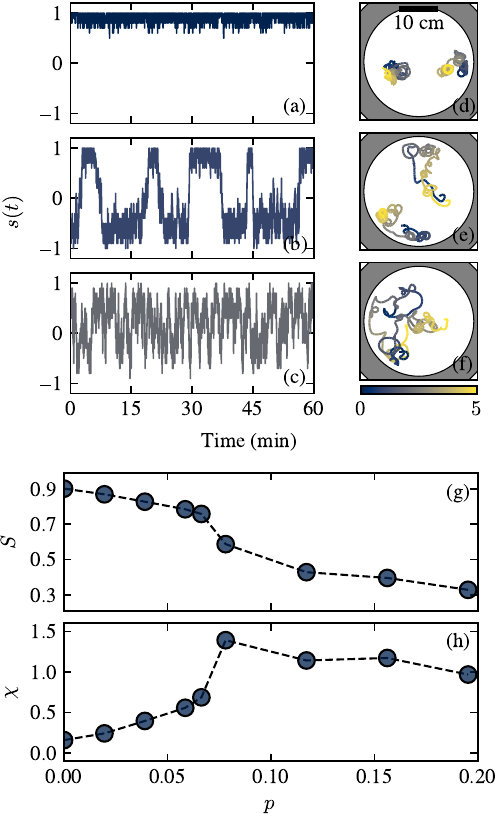}
\caption{\label{fig:experimental}Experimental results. (a)-(c) Temporal evolution of the order parameter for \(p = 0.02\), \(0.08\), and \(0.20\), respectively. (d)-(f) Trajectories performed by two robots during 5 minutes for the same values of \(p\) mentioned before. (g)-(h) Stationary average of the order parameter and susceptibility as a function of \(p\).
}
\end{figure}

Figure~\ref{fig:experimental} presents results for \(p = 0.02\), \(0.08\), and \(0.20\). It can be observed that in the case of the lowest \(p\), the system becomes almost completely ordered, showing some fluctuations due to the low probability of changing the rotation direction. In the case of the highest \(p\), the order parameter exhibits random behavior. Finally, an intermediate value of \(p\) shows that the order parameter fluctuates between the two possible ordered states. Furthermore, analyzing the individual trajectories of the particles, it was observed that they undergo a noticeable change in behavior. While for small values of \(p\) [Fig.~\ref{fig:experimental}(d)], the trajectories were mostly circular and localized—resembling the behavior of CAPs—increasing this parameter caused the trajectories to cover a larger area in the same time interval with a less defined structure, similar to that of ABPs [Fig.~\ref{fig:experimental}(e) and (f)].

Then, for each value of \(p\), we calculate \(S = \left\langle\left|s(t)\right|\right\rangle\) and \(\chi = N\left( \left\langle\left|s(t)\right|^2\right\rangle - \Bigl\langle\left|s(t)\right|\Bigr\rangle^2 \right)\), the temporal average of the order parameter and the susceptibility, respectively. \(S\) measures the order state throughout the experiment: a value close to $0$ implies disorder in the rotation directions, while a value close to $1$ indicates an ordered system regardless of the direction. The susceptibility measures the fluctuations of \(S\) during the experiment. Results are shown in Figs.~\ref{fig:experimental}(g) and (h), where a monotonic relationship between the order parameter and the value of \(p\) can be observed. Additionally, a change in concavity is noticeable between the values of \(p = 0.05\) and \(p = 0.10\), along with a maximum value of susceptibility in the same interval. These findings are consistent with an order transition that we will characterize in the following section.

\section{Numerical results}
\label{sec:num}

To delve deeper into the study of the system, a model was proposed that incorporates the main experimental features: polar circular particles that move by rotating in both directions and can exchange information with neighboring particles to update their motion state. The particles are characterized by their position \(\bm{r}\), orientation \(\bm{n} = (\cos{\theta},\sin{\theta})\), and motion state \(\sigma_i\). The evolutions of the position and orientation are given by:
\begin{align}
\dot{\bm{r}}_i & = v_0\ \bm{n} + F_{ij}\ , \\
\dot{\theta} & = \sigma_i(t) \frac{\left|\dot{\bm{r}}_i\right|}{R_p} + \eta\ , \\
F_{ij} & = -\kappa \epsilon \hat{\rho}_{ij}
\end{align}
where \(v_0\) is the translational velocity of the particle, \(R_p\) is the particle's radius, and \(\eta\) is an uncorrelated, zero-mean Gaussian perturbation with amplitude \(D\), accounting for the observed variability in rotation in the experiments. \(F_{ij}\) is the interaction force with other particles and the arena. This is a repulsive force depending on the overlap distance \(\epsilon\) between the two involved objects, an intensity constant \(\kappa\), and the normal contact direction given by \(\hat{\rho}_{ij} = \frac{\bm{r}_j-\bm{r}_i}{\left| \bm{r}_j-\bm{r}_i \right|}\). A free particle will move with a velocity \(v_0\), describing, on average, circular trajectories with a turning radius of \(R_p\) and a direction given by \(\sigma_i(t)\). This is motivated by the experimental system where the robot's rotations are around one of its rear legs. The update of the state \(\sigma_i\) followed the same rules as in the experimental system: for each particle, \(\Phi_i\) was calculated with a period \(T\), considering \(M=2\) particles within a radius of $10\ \mathrm{cm}$.

We simulated the equations using the Euler-Maruyama algorithm with an integration step of \(dt = 10^{-2}\ \mathrm{s}\) and a simulation length of \(10^5\ \mathrm{s}\). We varied the number of agents in the range \(N=\left[ 20-2000 \right]\) and the confinement radius \(R=\sqrt{\frac{N}{\pi \rho_0}}\) to fix the density at the value \(\rho_0 = 0.028\ \mathrm{cm^{-2}}\) used in the experiment.

\begin{figure}
\includegraphics{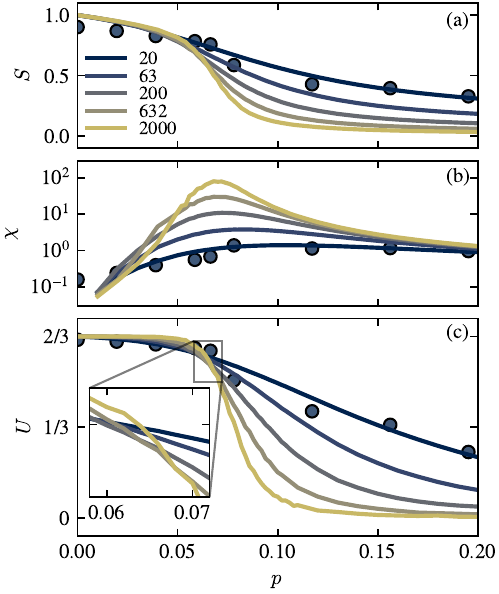}
\caption{\label{fig:numerical} Numerical results. (a) Order parameter, (b) susceptibility, and (c) Binder coefficient as a function of \(p\). The solid lines depict results for 5 system sizes, and the symbols represent experimental data for \(N=20\). The inset in (c) provides a detailed view of the intersection between the different curves. The parameters used in the simulation were: $v_0 = 0.825\ \mathrm{cm\ s}^{-1}$, $R_P = 1.65\ \mathrm{cm}$, $D = 0.05$, and $\kappa = 50\ \mathrm{s}^{-1}$.
}
\end{figure}

In Fig.~\ref{fig:numerical}, the results of the order parameter \(S\), susceptibility \(\chi\), and the reduced fourth-order cumulant of Binder defined as \(U = 1 - \frac{\left\langle s(t)^4\right\rangle}{3\left\langle s(t)^2\right\rangle^2} \) are shown. The results are the average of 100 independent realizations. The order parameter [Fig.~\ref{fig:numerical}(a)] shows that, for this particular density, when the value of \(p\) is reduced below a critical value, there is a transition from a disordered phase to an ordered one. Additionally, it can be observed that the transition is more pronounced when considering larger systems. Figure~\ref{fig:numerical}(b) shows the dependence of susceptibility on \(p\) and \(N\). The curves exhibit peak values whose location depends on \(N\). As the system size increases, the maximum value of \(\chi\) is located at smaller values \(p_\chi(N)\) and appears to saturate at a value \(p_c\). On the other hand, Fig.~\ref{fig:numerical}(c) shows the curves of the Binder parameter \(U(p)\). It can be observed that two curves corresponding to sizes \(N_i\) and \(N_j\) intersect at a probability value \(p_B(N_i,N_j)\) and, moreover, the location of the intersection appears to saturate at \(p_c\) when observing larger systems. The absence of negative values of the Binder parameter indicates that the order transition is consistent with a second-order one. Experimental results are also presented in the figure to demonstrate the excellent agreement with the numerical findings.

\begin{figure}
\includegraphics{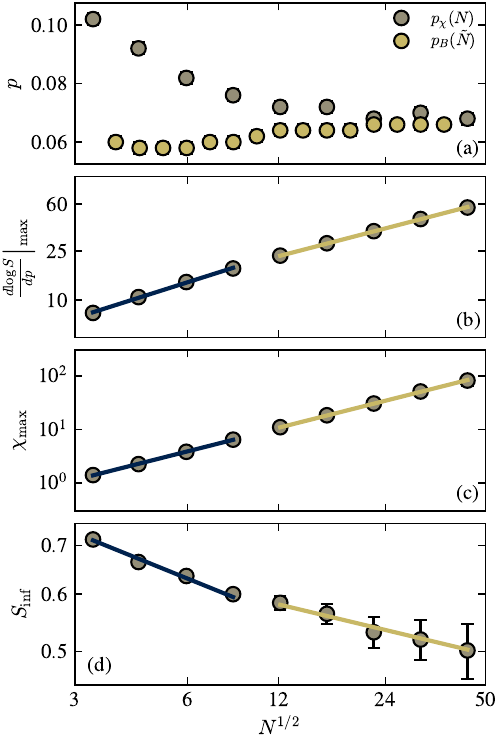}
\caption{\label{fig:scaling} (a) Positions of the maxima of the susceptibility and the intersections of the Binder parameter as a function of the system size ($N^{1/2}$) in a semilog scale. (b) Maximum value of the derivative of the logarithm of the order parameter. (c) Maximum values of the susceptibility. (d) Value of the order parameter evaluated at the inflection point. The solid lines are fits made in two ranges of system size. (b)-(d) are in log-log scale.
}
\end{figure}

Figure~\ref{fig:scaling}(a) shows the values of \(p_\chi\) and \(p_B\) as a function of the corresponding size. To describe the system's size, we used \(N^{1/2}\), which, at constant density, is proportional to the system size \(R\). In the case of \(p_B\), we used an equivalent value of \(N\) defined as \(\tilde{N}=\sqrt{N_i N_j}\). The results of \(p_B(\tilde{N})\) show a variation greater than 10\%, contrary to what the theory suggests, where all \(U\) curves should intersect at the critical value \(p_c\). However, when considering larger systems, the two datasets converge to a critical value estimated at \(p_c=0.067(1)\). Next, to characterize the transition, we studied the numerical results through a finite-size scaling analysis and calculated the standard critical exponents: \(\nu\), \(\gamma\), and \(\beta\). For the first, we studied the scaling behavior of the maximum value of the derivative of the logarithm of the order parameter, following a power-law of the form: \cite{lopez2009critical}
\begin{eqnarray}
\left. \frac{d\log{S}}{dp}  \right\vert_\mathrm{max} \propto N^{1/2\nu}\ .
\label{eq:nu}
\end{eqnarray}
In Fig.~\ref{fig:scaling}(b), we plotted the obtained values as a function of the system size on a log-log scale. We found that the data does not follow a linear trend across the entire range. Therefore, we performed two fits of Eq.~\eqref{eq:nu}: one for the lower values of the system size and another for the higher values. For the first one, the calibration led us to a value $\nu^L = 1.04(3)$, and for the second, we obtained $\nu^H = 1.27(1)$. The results show significant differences, indicating a change in the behavior of the correlation length with increasing system size.

Then, we study the relationship between the maximum susceptibility value \(\chi_\mathrm{max}\) and the characteristic system size \(N\). For this, the scaling law is expressed as follows:
\begin{eqnarray}
\chi_\mathrm{max} \propto N^{-\gamma/2\nu}\ .
\label{eq:gnu}
\end{eqnarray}
Encouraged by the previous results, we also performed two fits of Eq.~\eqref{eq:gnu}. In Fig.~\ref{fig:scaling}(c), the obtained results are shown. For the lower values, we obtained \((\gamma/\nu)^L = 1.77(2)\), while for the higher, the value was \((\gamma/\nu)^H = 1.76(1)\). In the case of susceptibility, we found that the results between both regions do not present significant differences, and moreover, the value \(\gamma/\nu\) corresponds to that of the 2D Ising model.

Finally, we investigate the scaling behavior of the order parameter at the inflection point \(S_\mathrm{inf}\) in relation to $N^{1/2}$. This scaling behavior is described by the following power-law relationship:
\begin{eqnarray}
S_\mathrm{inf} \propto N^{-\beta/2\nu}\ .
\end{eqnarray}
Figure~\ref{fig:scaling}(d) shows the obtained results. The inflection point of each curve was determined by finding the value of \(p\) corresponding to the maximum of the numerical derivative of \(S\). We performed two fits again and obtained \((\beta/\nu)^L=0.21(2)\) and \((\beta/\nu)^H=0.123(8)\), respectively. The results differ significantly, and, similar to the susceptibility, in the higher range of sizes, the value of \(\beta/\nu\) corresponds to that of the Ising model in 2D.

\begin{figure}
\includegraphics{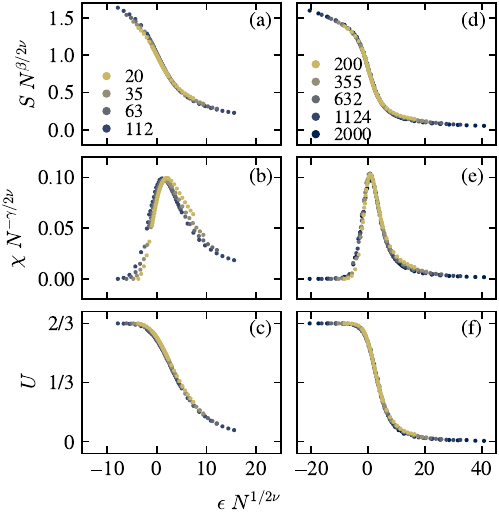}
\caption{\label{fig:collapse}Collapse of the data considering (a)-(c) the smaller sizes, and (d)-(e) the larger ones. The figures were plotted using $p_c = 0.067$ and the corresponding values of $\nu^{L,H}$, $(\gamma/\nu)^{L,H}$, and $(\beta/\nu)^{L,H}$ as indicated in the text.
}
\end{figure}

The finite-size scaling behavior can be verified by plotting \(S\ N^{\beta/2\nu}\), \(\chi\ N^{-\gamma/2\nu}\), and \(U\) as a function of \(\epsilon\ N^{1/2\nu}\), where \(\epsilon = p/p_c-1\) and \(p_c = 0.067\). In Figs.~\ref{fig:collapse}(a)-(c), the collapse of the curves for the smaller sizes is shown. While the collapse seems relatively appropriate for the order and Binder parameters, a strong discrepancy is observed for the susceptibility. On the other hand, for the larger sizes [Figs.~\ref{fig:collapse}(d)-(f)], the collapse of all three quantities is satisfactory. Furthermore, we found that the hyperscaling relation $2\beta/\nu+\gamma/\nu=2$ holds in the higher range, resulting in $2.01(2)$, but it fails in the lower one, yielding $2.18(4)$.

The experimental results in Fig.~\ref{fig:experimental}(d)-(f) showed that the type of trajectories performed by the particles is influenced by the value of $p$. To characterize this behavior, we calculated the mean square angular displacement (MSAD) given by $\mathrm{MSAD}=\left\langle \Delta \theta (t)^{2}\right\rangle$, where the brackets denote the average over all particles. We chose to perform the calculation for the case of $N = 2000$ to have the largest statistical ensemble.

\begin{figure}
\includegraphics{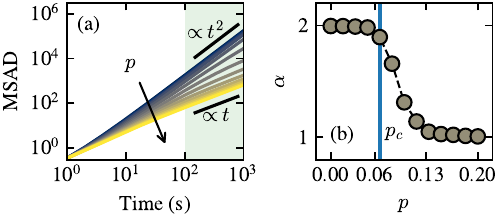}
\caption{\label{fig:msad} (a) Mean squared angular displacement for different values of $p$. The results correspond to the $N=2000$ system. The arrow indicates the direction of increasing $p$. The black lines show data trend. The shaded region represents the set of data fitted by a function of the form $t^\alpha$. (b) Values of $\alpha$ obtained from the previous fits. The vertical line indicates the estimated value of $p_c$ from Fig.~\ref{fig:scaling}(a).
}
\end{figure}

Figure~\ref{fig:msad}(a) shows the results we obtained for different values of $p$. We analyzed the MSAD results on a log-log scale, where a temporal evolution with a slope of 2 indicates circular trajectories (CAP). On the other hand, a slope of 1 implies that the particle orientations and their trajectories follow symmetric diffusive behavior (ABP). The results show that at low values of $p$, i.e., during the ordered phase, the particles exhibit a chiral behavior. As $p$ increases and the system enters the disordered phase, the particles change their direction of motion randomly, following Brownian trajectories. To quantify this change, we performed fits of the MSAD in the stationary range $t>100\ \mathrm{s}$ using functions of the form $t^\alpha$. In Fig.~\ref{fig:msad}(b), we present the values of $\alpha$ obtained for each $p$. It can be observed that there is a change in the particle motion behavior upon crossing the critical value $p_c$: the phase transition from ordered to disordered state results in a transition from CAP to ABP behavior.

Finally, we characterize the density of particles within the arena as a function of $p$. We divide the space into two regions - one bulk and the other surface - using a circle of radius $R_c = R-2R_P$. We calculate the density of bulk particles according to:
\begin{eqnarray}
\rho_B(R) = \frac{N_B}{\pi R_c^2}\ ,
\end{eqnarray}
where $N_B$ is the number of particles located at a distance from the center of the arena less than $R_c$. We also calculate the density of particles on the surface using:
\begin{eqnarray}
\rho_S(R) = \frac{N-N_B}{\pi \left( R^2-R_c^2 \right) }\ . 
\end{eqnarray}

\begin{figure}
\includegraphics{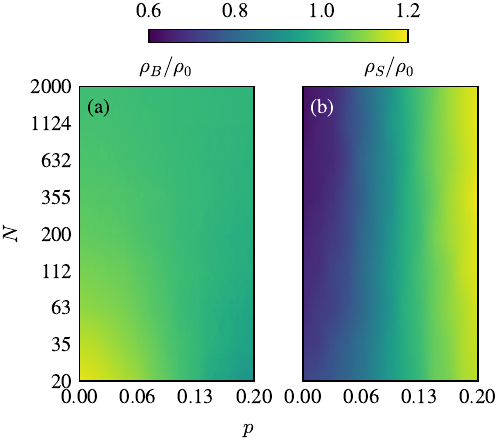}
\caption{\label{fig:density} Relative density of (a) bulk and (b) surface as a function of $p$ and $N$.
}
\end{figure}

Figure~\ref{fig:density}(a) shows the density results obtained for the bulk normalized by the reference density value $\rho_0$. It can be observed that for smaller system sizes, $\rho_B$ is affected by the value of $p$. For example, in the case of $N=20$, the density ratio decreases with an increase in the control parameter. This implies a noticeable migration of particles toward the surface region. When analyzing larger sizes, we observed that the density becomes uniform across the entire range of $p$. Subsequently, in Fig.~\ref{fig:density}(b), we present the results related to the density of particles near the surface. Across all the analyzed size range, it can be observed that increasing the $p$ enhances the particle density at the surface. This result does not surprise us, as the accumulation of particles at the arena boundaries is a typical characteristic of active Brownian particles \cite{bechinger2016active}. This explains why the bulk density is reduced at higher values of $p$, where the particles in the system behave like ABPs. Additionally, it is worth noting that the number of agents that can be found on the surface scales with $N^{1/2}$, while in the bulk, it grows as $N$. For this reason, the average density in the bulk tends to $\rho_0$ and independent of \(p\) as the system size increases. Finally, the density variation in smaller systems could impact local interactions and lead to size-sensitive effects, as observed in the scaling analysis.

\section{Conclusion}
\label{sec:con}

In summary, we investigated the emergent behavior of an interacting particle system confined within a circular arena. For this purpose, we employed 20 commercially available robots known as Kilobots. These robots possess differential locomotion capabilities and can communicate with their neighbors. They were programmed with two rotational motion states: one in the clockwise and the other in the counterclockwise direction. Using a stochastic interaction that considered the rotational states of two neighboring units, each robot had the ability to change its own rotational state. The interaction was mediated by a control parameter, where higher values led to a disordered and symmetric behavior, implying that robots rotated randomly in both directions. Upon reducing the parameter below a certain threshold, the system exhibited a symmetry breaking where all robots rotated in a uniform direction. Furthermore, we observed that as a consequence of this transition, the trajectories of the particles exhibited a noticeable change: below the threshold, they were circular, while above it, they displayed random motion.

Then, we employed a mathematical model to extend the results to larger systems. This enabled us to characterize the transition of the system using the order parameter, susceptibility, and Binder parameter. We found that the results did not meet the scaling relation in the studied range. Consequently, we opted to conduct a finite-size analysis in two regions. The critical exponents determined for the smaller size region failed to satisfy the hyperscaling relationship and resulted in poor curve collapses, notably in the susceptibility. Conversely, within the larger size region, the exponents adhered to the hyperscaling relationship, and the curves displayed excellent collapse in scaling renormalization. An intriguing aspect of the critical exponents in the higher region is that both $\gamma/\nu=1.76(1)$ and $\beta/\nu=0.123(8)$ agree with those of the 2D Ising model. However, the value $\nu=1.27(1)$ significantly differs. In addition, we found that the transition type corresponded to that of a second order.  

We also studied the mean square angular displacement (MASD) and confirmed what was observed in the experiments: the system's order transition is accompanied by a transition in the particles' trajectories. While in the disordered state, the particles exhibit diffusive trajectories characterized by an evolution like $\mathrm{MASD}\propto t$, in the ordered state, we found a relationship of the form $\mathrm{MASD}\propto t^2$, indicating chiral behavior.

Finally, we studied the bulk density as a function of the control parameter and system size. We observed that in smaller systems, the bulk density is notably sensitive to the control parameter. This is due to the fact that for higher values of $p$, the particles behave like ABPs and tend to accumulate at the boundaries of the arena, as shown by the surface density calculations. However, in the larger systems, the bulk density shows independence from the control parameter. This is attributed to the limitation on the number of particles that can reside on the surface, which scales proportionally with the system's radius, that is, $N^{1/2}$.

The findings from this study hold significant implications in the field of swarm intelligence, wherein system functionalities often scale with the number of particles without requiring a redefinition of interaction types. Specifically, the behavior demonstrated by this system can be adopted as a search strategy, where the control parameter acts as the response to a environmental stimulus.

\section*{ACKNOWLEDGMENTS}
This work was funded by project PICT 2019-00511 (Agencia Nacional de Promoci\'on Cient\'ifica y Tecnol\'ogica, Argentina).

\bibliography{manusBib}

\end{document}